\documentclass{elsart}
\usepackage{psfig}
\usepackage{natbib}
\def\Fvar{\ifmmode F_{\rm var} \else $F_{\rm var}$\fi}
\def\Rmax{\ifmmode R_{\rm max} \else $R_{\rm max}$\fi}
\def\tcent{\ifmmode \tau_{\rm cent} \else $\tau_{\rm cent}$\fi}

\def\kms{\ifmmode {\rm km\ s}^{-1} \else km s$^{-1}$\fi}
\def\Msun{\ifmmode M_{\odot} \else $M_{\odot}$\fi}
\def\Lsun{\ifmmode L_{\odot} \else $L_{\odot}$\fi}
\def\Rs{\ifmmode R_{\rm S} \else $R_{\rm S}$\fi}
\def\qo{\ifmmode q_{\rm o} \else $q_{\rm o}$\fi}
\def\Ho{\ifmmode H_{\rm o} \else $H_{\rm o}$\fi}
\def\ho{\ifmmode h_{\rm o} \else $h_{\rm o}$\fi}
\def\ltsim{\raisebox{-.5ex}{$\;\stackrel{<}{\sim}\;$}}

\def\vFWHM{\ifmmode V_{\mbox{\tiny FWHM}} \else
            $V_{\mbox{\tiny FWHM}}$\fi}
\def\CCF{\ifmmode F_{\it CCF} \else $F_{\it CCF}$\fi}
\def\ACF{\ifmmode F_{\it ACF} \else $F_{\it ACF}$\fi}
\def\Halpha{\ifmmode {\rm H}\alpha \else H$\alpha$\fi}
\def\Hbeta{\ifmmode {\rm H}\beta \else H$\beta$\fi}
\def\4686{\ifmmode {\mbox{\rm He\,}{\sc ii}}\,\lambda4686
	\else He\,{\sc ii}\,$\lambda4686$\fi}
\def\Hgamma{\ifmmode {\rm H}\gamma \else H$\gamma$\fi}
\def\Hdelta{\ifmmode {\rm H}\delta \else H$\delta$\fi}
\def\Lya{\ifmmode {\rm Ly}\alpha \else Ly$\alpha$\fi}
\def\Lyb{\ifmmode {\rm Ly}\beta \else Ly$\beta$\fi}

\def\ciii{\ifmmode {\rm C}\,{\sc iii} \else C\,{\sc iii}\fi}
\def\civ{\ifmmode {\rm C}\,{\sc iv} \else C\,{\sc iv}\fi}

\def\o5007{[O\,{\sc iii}]\,$\lambda5007$}

\begin{document}
\runauthor{Peterson, McHardy, and Wilkes}
\begin{frontmatter}
\title{Variability of NGC~4051
and the Nature of Narrow-Line Seyfert 1 Galaxies}
\author[OSU]{Bradley M.\ Peterson}
\author[Southampton]{Ian M.\ McHardy}
\author[CfA]{Belinda J.\ Wilkes}
\address[OSU]{Department of Astronomy, The Ohio State University}
\address[Southampton]{Department of Physics and Astronomy,
University of Southampton}
\address[CfA]{Harvard-Smithsonian Center for Astrophysics}
\begin{abstract}
We report on a three-year program of coordinated
X-ray and optical monitoring of the narrow-line Seyfert 1 galaxy
NGC~4051. The rapid continuum variations observed in the X-ray
spectra are not detected in the optical, although the X-ray and
optical continuum fluxes are correlated on time scales of many weeks
and longer. Variations in the flux of the broad \Hbeta\ line
are found to lag behind the optical continuum variations by
approximately 6 days (with an uncertainty of 2--3 days), and combining
this with the line width yields a virial mass estimate of
$\sim1.1 \times 10^6$\,\Msun, at the very low end of the distribution
of AGN masses measured by line reverberation. Strong variability of
\4686\ is also detected, and the response time measured is similar to
that of \Hbeta, but with a much larger uncertainty. The
\4686\ line is almost five times broader than \Hbeta, and it is
strongly blueward asymmetric, as are the high-ionization UV lines
recorded in archive spectra of NGC~4051. The data are consistent
with the Balmer lines arising in a low-inclination (nearly face-on)
disk-like configuration, and the high-ionization lines arising in
an outflowing wind, of which we observe preferentially the near side.
During the third year of monitoring,
both the X-ray continuum and the \4686\ line went into extremely
low states, although the optical continuum and the \Hbeta\ broad line
were both still present and variable. We suggest that the inner
part of the accretion disk may have gone into an advection-dominated
state, yielding little radiation from the hotter inner disk.
\end{abstract}
\end{frontmatter}

\section{Introduction}
Beginning in early 1996, the International AGN Watch\footnote{All
International AGN Watch papers and data are available at the AGN
Watch website at URL
{\sf http://www.astronomy.ohio-state.edu/$\sim$agnwatch/}.}
undertook a program of 
contemporaneous X-ray and optical spectroscopic
monitoring of the galaxy NGC~4051, the only NLS1
galaxy in Seyfert's \cite{Sey43} original list of high surface-brightness
galaxies with strong emission lines. The X-ray variability
characteristics of NGC~4051 are typical of the NLS1 class
\cite{Law87,McH95}.
X-ray observations were made 
with the {\em Rossi X-Ray Timing Explorer (RXTE)},
and optical spectra were obtained with 
the Perkins 1.8-m telescope at Lowell Observatory
and the 1.6-m Tillinghast Reflector on Mt.\ Hopkins.
The purpose of this program has been twofold:
\begin{enumerate}
\item To determine the nature of the relationship between the
X-ray and UV--optical continuum variations. This is
a particularly interesting question in the case of NGC~4051
since the X-ray flux dropped to an extremely low level
towards the end of this campaign \cite{Utt99}. 
\item To determine the size of the broad-line region (BLR) and 
virial mass of the central object via
reverberation techniques \cite{Bla82,Pet93}.
\end{enumerate}

The emission-line cross-correlation lag $\tau$
can be taken to be the light-travel time across the BLR, so
the BLR size is given by $r=c\tau$.
By combining this with the emission-line width $\vFWHM$,
the mass of the central source can be inferred to be
\begin{equation}
M = \frac{f \vFWHM^2 c\tau}{G},
\end{equation}
where $f$ is a factor of order unity that depends on the still
unknown geometry and kinematics of the BLR. There is an
implicit assumption that the gravitational force of the
central object dominates the kinematics of the BLR; this is
formally unproven, but at least in the case of the well-studied
Seyfert 1 galaxy NGC~5548, the reverberation-mapping data are
consistent with the required $\vFWHM \propto r^{-1/2}$ relationship
\cite{Pet99}. Virial mass estimates based on
reverberation-mapping data are now available
for nearly 40 AGNs \cite{Wan99,Kas00}.

\section{Results}
In Fig.\ 1, we show the continuum and emission-line light
curves obtained during this three-year program.
While there is a clear lack of correlated short time-scale
behavior of the X-ray and optical continua, the light curves
in Fig.\ 1 suggest that a correlation on longer time scales
is possible. To test this quantitatively, we have 
suppressed the rapid variations, smoothing both the optical continuum
and X-ray light curves with a rectangular function
of width 30 days, 
similar to what was done 
in a comparison of X-ray and optical variability in
the Seyfert 1 galaxy NGC~3516 \cite{Mao00}.  Cross-correlation of the
overlapping parts of these light curves
yields a lag of the optical variations relative to the
X-ray of $\tau= 6^{+62}_{-112}$ days with a correlation
coefficient $r_{\rm max} = 0.74$, i.e., the mean X-ray and optical
fluxes are indeed  correlated once the high-frequency variability
is suppressed. The lag between variations in the two wavebands 
is highly uncertain, but consistent with zero or any small
time lag expected in the continuum emitting region.

\begin{figure}
\mbox{
\psfig{figure=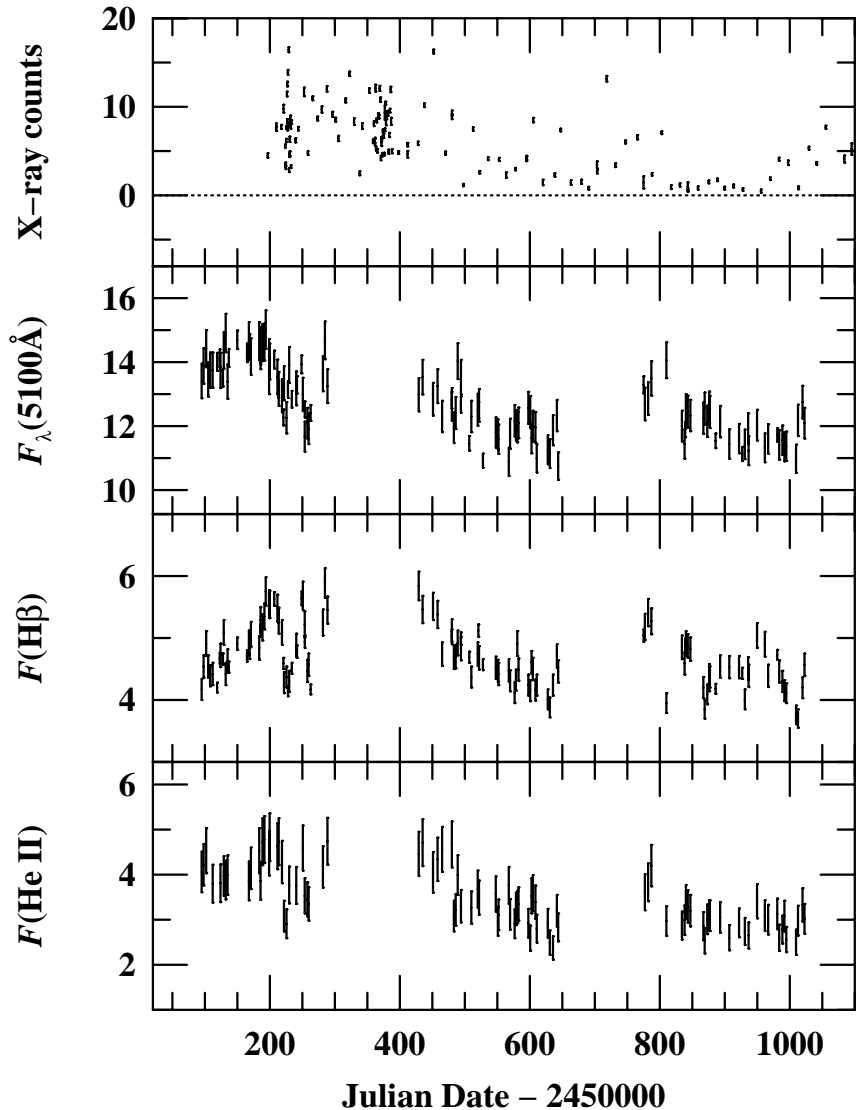,width=12cm,angle=0}}
\caption{Continuum and emission-line light curves of NGC~4051 between
January 1996 and June 1998. Top panel is 2--10\,keV count rate.
Second panel is optical continuum flux in units of 
$10^{-15}$ ergs s$^{-1}$ cm$^{-2}$ \AA$^{-1}$.
Third and fourth panels are \Hbeta\ and \4686\ fluxes, respectively, in
units of $10^{-13}$ ergs s$^{-1}$ cm$^{-2}$.}
\end{figure}

The time delay between 
continuum and emission-line variations has been determined by
cross-correlation of the light curves. The period from
JD2450183 to JD2450262 is particularly well-suited to this
type of analysis because the 
light curves are well-sampled and the character of the variations
permits an accurate cross-correlation measurement. 
The time delays, or lags, we measure are
$\tau = 5.9^{+3.1}_{-2.0}$ days for \Hbeta\ and
$\tau = 4.5^{+4.9}_{-5.6}$ days for \4686, where the
uncertainties are based on a model-independent Monte Carlo
method \cite{Pet98}.

By combining this lag with the Doppler width of the emission
line, we can estimate a virial mass, as in eq.\ (1);
for consistency with previous work \cite{Wan99, Kas00},
we take $f=3/\sqrt{2}$.
Since the broad emission-line features are comprised of
a number of different components (or contaminants), it
is desirable to measure the Doppler width of only
the {\em variable} part of the emission line.
In order to isolate the 
variable part of the emission line and exclude constant
components (such as contamination from the narrow-line region),
we measure the relevant line widths in the 
root-mean-square (rms)
spectrum constructed from the spectra used to determine the
time lag; the principal virtue of the rms spectrum is that
it isolates the features in the spectrum that are actually varying.
We measure widths $\vFWHM = 1110 \pm 190$\,\kms\ for \Hbeta,
and $\vFWHM = 5430 \pm 510$\,\kms\ for \4686; the latter is
also strongly blueward asymmetric. Examination of archival
{\it IUE}\ spectra show that the UV high-ionization lines
are similarly asymmetric. On the basis of 
the \Hbeta\ variations, a virial
mass of $1.1^{+0.8}_{-0.5} \times 10^6\,\Msun$
is inferred; unfortunately however, the extremely large uncertainty 
in the \4686\ lag renders the virial mass obtained from it
not very enlightening
(i.e., $M = 19.4^{+21.5}_{-24.4} \times 10^6\,\Msun$), 
but it is consistent with the \Hbeta\ result.
It is also important to keep in mind that because of the unknown
geometry and kinematics of the BLR, the virial mass is reliable
to only about an order of magnitude, i.e., the systematic uncertainties
are much larger than the errors quoted here.

\section{Discussion}

\subsection{The Continuum}

It has already been pointed out based on these same {\em RXTE} data 
\cite{Utt99} that the X-ray continuum
of NGC~4051 virtually ``turned off'' in early 1998. 
However, the optical spectroscopic
data show that the optical continuum and emission lines
(and therefore, by inference, the ionizing UV continuum)
did not disappear at the same time. 
Figure 2a shows the {\em RXTE} light curve.
The two lower panels of Fig.\ 2 show the optical rms spectra based
on data obtained during the intervals marked on Fig.\ 2a.
Figure 2b shows the rms spectrum obtained during
a period of relatively active X-ray and optical continuum variability.
The rms spectrum in Fig.\ 2c is based on spectra obtained
during the period when the X-rays were about an order of magnitude
fainter than normal. Figure 2b shows that
when the X-ray flux was high and variable, the optical continuum
and the \Hbeta\ and \4686\ emission lines were also variable.
Figure 2c, however, shows that when the
X-rays were in a faint state, both the optical continuum and
the \Hbeta\ line were still varying, but the \4686\
emission line was no longer varying.  A possible interpretation of 
the behavior of NGC~4051 is that the inner X-ray producing
part of the accretion disk has entered
an advection-dominated accretion-flow (ADAF) state, in
which radiation is emitted with very low efficiency
\cite{Nar94,Nar98}. The outer
part of the disk, which produces the UV--optical continuum and
drives the emission lines, remains relatively unaffected, however.
The implication is that there is a transition radius
inside of which the disk is an ADAF and
outside of which it radiates efficiently, perhaps like
a classical thin disk \cite{Sha73}, and the
persistence of the optical continuum and the \Hbeta\ emission line suggests
that this transition radius is somewhere between the
regions that are most responsible for the soft X-rays and
the H-ionizing continuum. 

\begin{figure}
\psfig{figure=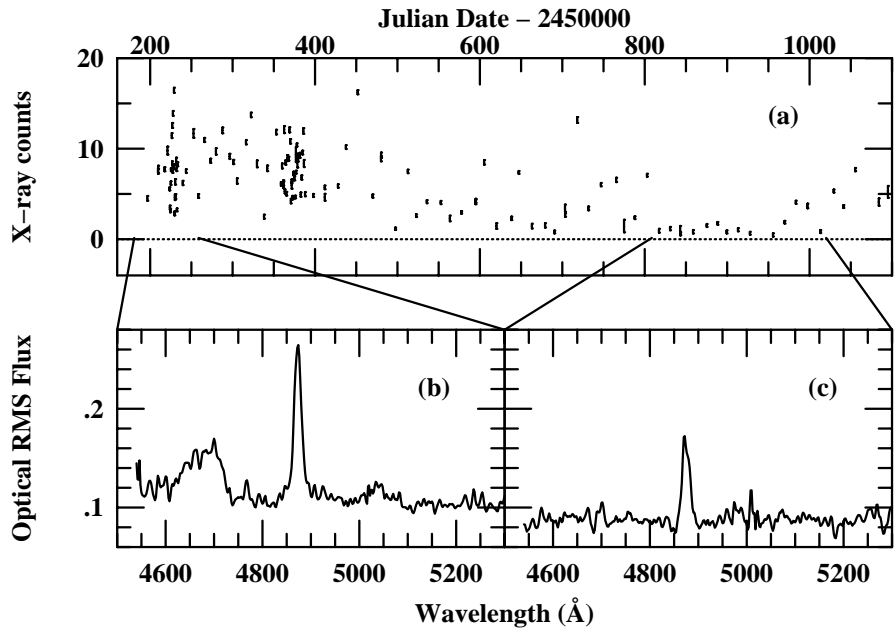,width=12cm,angle=0}
\caption{Top panel shows the X-ray light curve as in Fig.\ 1.
Bottom left panel shows the optical rms spectrum based on
data obtained between JD2450183 and JD2450262.
Bottom right panel shows the optical rms spectrum based on
data obtained between JD2450810 and JD2451022.}
\end{figure}

\subsection{The Virial Mass and Implications for NLS1s}

Reverberation-based size estimates for the
broad emission lines and resulting virial mass estimates
provide a potential means of distinguishing among the various
NLS1 models. In Fig.\ 3a, we show the relationship between
the BLR radius as measured from the Balmer-line lags as a function
of the optical continuum luminosity for all AGNs with 
Balmer-line lags known to reasonable accuracy \cite{Wan99, Kas00}.
This compilation contains six
additional AGNs that could be classified as NLS1s
as they meet the criterion $\vFWHM \ltsim 2000$\,\kms. 
The best-fit regression line ($R_{\rm BLR} \propto L^{0.62 \pm 0.02}$),
based on all objects (excluding NGC~4051), is shown as a dotted line.
NGC~4051 lies approximately 2.8$\sigma$ above this regression
line, although all the other narrow-line objects clearly fall
in the locus defined by the AGNs with broader lines. 
Given the large dispersion in this relationship, reflected
in the high $\chi^2_{\nu} (= 15.7)$ of the fit, the offset
of NGC~4051 is not unusual. 

\begin{figure}
\psfig{figure=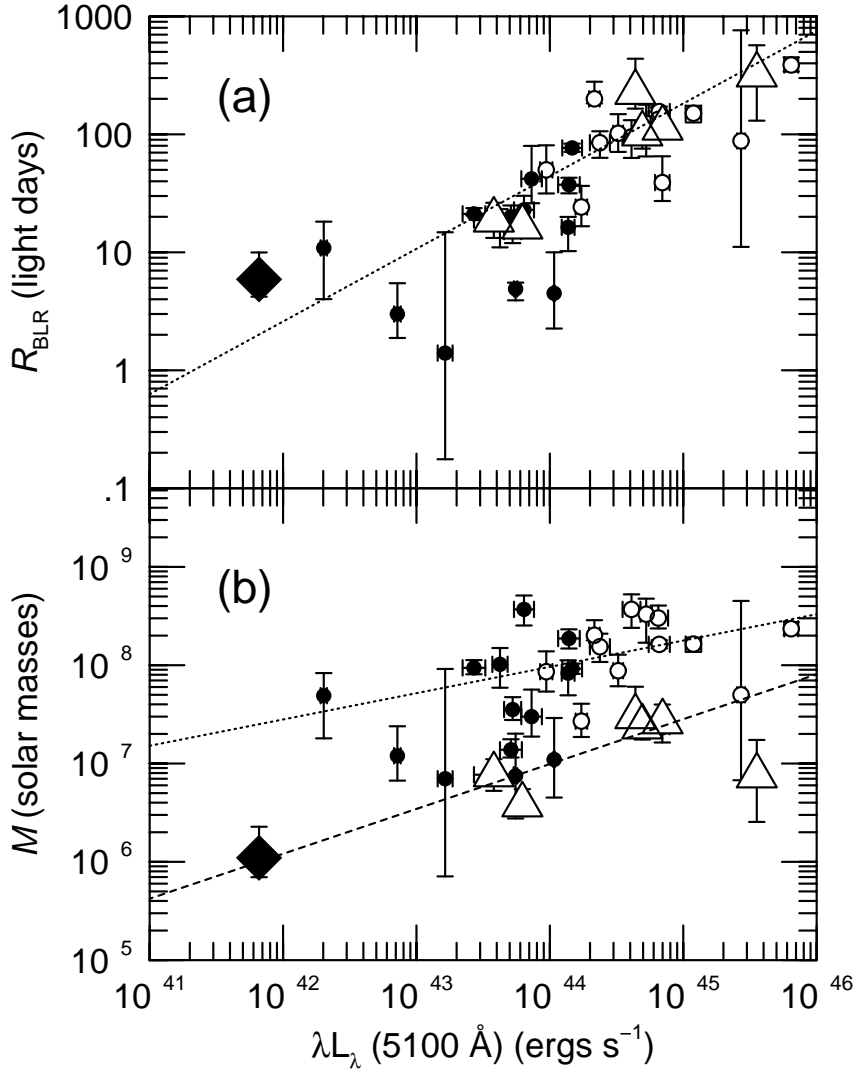,width=12cm,angle=0}
\caption{(a). The relationship between size of the Balmer-line emitting
region and optical luminosity. The dotted line is a least-squares
fit to all the data, excluding NGC~4051.
(b). The relationship between virial mass and optical luminosity.
The dashed  line is a least-squares fit to the narrow-line objects
only, and the dotted line is a least-squares fit to all of the
other points. In both panels, the filled circles are Seyfert
1 galaxies and the open circles are QSOs.
The large triangles are those AGNs in which \Hbeta\ has a half
width less than 2000\,\kms. NGC~4051 is shown as a filled diamond.}
\end{figure}

Figure 3b shows the mass--luminosity relationship for these
AGNs. We show (a) the best-fit regression line
based on all objects {\em except} the seven narrow-line objects
and (b) that based on the narrow-line objects alone. 
These two fits are separated by typically an order of magnitude
in black-hole mass; the black holes in the narrow-line objects 
are about a factor of 10 lower than those of other AGNs of comparable
luminosity.

How well do these results allow us to distinguish among the
various explanations for the NLS1 phenomenon? We consider
the possibilities:
\begin{enumerate}
\item {\em Do the BLRs of NLS1s have anomalously large radii?}
The position of NGC~4051 in Fig.\ 3a might suggest that this is possible,
but the distribution of other narrow-line objects does not 
support this. Furthermore, as noted above, the scatter in
the BLR-radius luminosity relationship is very large, and 
NGC~4051 is in a statistical sense not the largest outlier in
this relationship (simply because other sources have smaller
uncertainties in their measured lags). 

\item {\em Are NLS1s simply low-inclination systems?}
If the BLR is a flattened system, at low inclination
(i.e., nearly face-on) the line widths will be decreased
by a factor $\sin i$, but the measured emission-line lags
will be relatively unaffected. On the other hand, assuming
that the UV--optical continuum arises in an accretion
disk at the same inclination, the apparent UV--optical luminosity
is higher at lower inclination. Thus, relative to similar sources at
intermediate inclination, the masses of low-inclination
sources will be underestimated, and their luminosities will
be overestimated, displacing the narrow-line objects in
Fig.\ 3b towards the lower right. This is generally consistent
with the location of all seven of the narrow-line objects,
including NGC~4051. The line transfer function for \Hbeta\
would provide a  more definitive test of this hypothesis
since it would allow determination of the inclination of the system.
This would require more and better data than we have
obtained in this experiment.

\item {\em Are NLS1s undermassive systems with relatively 
high accretion rates?} Again, the distribution of the
narrow-line objects, including NGC~4051, in Fig.\ 3b is consistent
with this hypothesis. The narrow-line sources on this plot
lie below the mass-luminosity relationship for other AGNs,
at the lower end of the envelope around this relationship.

\end{enumerate}

In summary, the hypothesis that NLS1s have unusually distant BLRs
for their luminosity
is probably not viable in general, although it could apply to
the specific case of NGC~4051. At the present time, however,
we cannot distinguish between the low-inclination and
low-mass, high accretion-rate hypotheses on the basis of the
reverberation results alone,  although considerations
based on X-ray data favor the latter. Indeed, it is entirely
possible that both effects (i.e., low inclination and low
black-hole mass) contribute. An improvement in 
the quality and quantity of the optical spectroscopic data 
could allow determination of the \Hbeta\ transfer function,
which could allow discrimination between these competing models.

The differences between the characteristics of the \Hbeta\ emission line
on one hand and of the high-ionization lines such as \4686\
on the other suggests a
two-component BLR, which has been proposed on numerous occasions on
other grounds \cite{Col88}.
In this particular case, an interpretation that is at least qualitatively
consistent with all the data and relatively simple
is that the Balmer lines arise primarily
in material that is in a flattened disk-like configuration at a low
inclination (to account for the narrow width of the \Hbeta\ line),
and the high-ionization lines arise in an outflowing polar wind,
of which we see preferentially the component on the near side of the disk
(to account for high velocity and blueward asymmetry) \cite{Col88}.
If indeed the \Hbeta\ emission arises primarily in a low-inclination
disk,  our virial mass estimate of $M=1.1\times10^6$\,\Msun\
might seriously underestimate the black-hole mass. It could then
be inferred that 
the NLS1 class might be best explained as low-inclination rather than
low-mass, high accretion-rate systems. 
The strong rapid X-ray variability of NLS1s 
seem to favor the latter explanation, but we note that
at least some low-inclination accretion-disk models predict relatively
strong, variable EUV/soft X-ray fluxes \cite{Net87, Mad88},
consistent with observations of NLS1s.

\section{Summary}
On the basis of three years of combined X-ray and optical spectroscopic
monitoring of the narrow-line Seyfert 1 galaxy NGC 4051, we
reach the following conclusions:
\begin{enumerate}
\item The rapid and strong X-ray variations that characterize narrow-line
Sey\-fert 1 galaxies are detected in our X-ray observations of
NGC~4051, but are not detected in the optical, consistent with
previous findings \cite{Don90}.
\item On time scales of many weeks and longer, there does appear to be
a correlation between the X-ray and optical continuum fluxes.
\item The variable part of the \Hbeta\ emission line has a Doppler
width of about $1100$\,\kms, and a time delay relative to the continuum
of about six days. Combining these quantities leads to a virial mass
estimate of approximately $1.1\times10^6$\,\Msun.
\item The \4686\ emission line is strongly variable, although an accurate
time delay cannot be measured. This line is about five times as broad
as the \Hbeta\ line, and is strongly blueward asymmetric, as are the
UV high-ionization lines in this object.
\item In the BLR radius--luminosity relationship, we find that narrow-line
objects (those with $\vFWHM \ltsim 2000$\,\kms) seem to fall on the same
locus as AGNs with broad lines.
\item In the virial mass--luminosity relationship, narrow-line objects
populate the low-mass end of a rather broad envelope; they have
virial masses typically an order of magnitude lower than other AGNs of
similar luminosity.
\item During the third year of this program, both the X-rays and
\4686\ nearly disappeared, while the optical continuum and broad \Hbeta\
emission line were only slightly fainter than previously and continued
to vary significantly. This suggests that the innermost part of the
accretion disk went into an ADAF state, greatly reducing the production
of high-energy continuum photons from the inner part of the accretion
disk. 
\end{enumerate}

A picture that is consistent with the emission-line characteristics is
one in which the Balmer lines arise primarily in a disk-like configuration seen
at low inclination (i.e., close to face-on), 
and the high-ionization lines arise 
primarily in an outflowing polar
wind. The high-ionization lines are blueward asymmetric because we see 
emission preferentially from the near side, with the far side 
at least partially obscured by the
disk component (which might be an extension of the accretion disk).
If this interpretation is correct, the virial mass estimate 
of $M=1.1\times10^6$\,\Msun\ is lower than the actual black hole mass
by a factor of $\sin i$. If NGC 4051 is typical of
the NLS1 class, then it might be that NLS1s are best described 
as  low-inclination rather than low-mass, high accretion-rate systems. 
While the observed X-ray characteristics of NLS1s
rather favor the latter explanation, they are not
inconsistent with an inclination-based origin. Indeed, the
full explanation of the phenomenon may involve {\em both}
inclination and black-hole mass.

\bigskip
For support of this work, we are grateful to 
the National Science Foundation (grant AST--9420080 
to The Ohio State University).


\end{document}